# Artificial intelligence enables mobile soil analysis for sustainable agriculture

Ademir Ferreira da Silva[1], Ricardo Luis Ohta[2], Jaione Tirapu Azpiroz[1], Matheus Esteves Fereira[1], Daniel Vitor Marçal[1], André Botelho[2], Tulio Coppola[2], Allysson Flavio Melo de Oliveira[2], Murilo Bettarello[3], Lauren Schneider[3], Rodrigo Vilaça[4], Noorunisha Abdool[5], Vanderlei Junior[6], Wellington Furlaneti[6], Pedro Augusto Malanga[6], Mathias Steiner[1,*]

[1] *IBM RESEARCH, Rio de Janeiro, RJ, Brazil*

[2] *IBM RESEARCH, São Paulo, SP, Brazil*

[3] *ENVERITAS, New York, NY, USA*

[4] *CSEM BRASIL, Belo Horizonte, MG, Brazil*

[5] *OMNIA FERTILIZERS RSA, Bryanston, Sandton, South Africa*

[6] *INTEGRADA, Londrina, PR, Brazil*

[*]mathiast@br.ibm.com

**For optimizing production yield while limiting negative environmental impact, sustainable agriculture benefits greatly from real-time, on-the-spot analysis of soil at low cost. Colorimetric paper sensors are ideal candidates for cheap and rapid chemical spot testing. However, their field application requires previously unattained paper sensor reliability and automated readout and analysis by means of integrated mobile communication, artificial intelligence, and cloud computing technologies. Here, we report such a mobile chemical**

**analysis system based on colorimetric paper sensors that operates under tropical field conditions. By mapping topsoil pH in a field with an area of 9 hectares, we have benchmarked the mobile system against precision agriculture standards following a protocol with reference analysis of compound soil samples. As compared with routine lab analysis, our mobile soil analysis system has correctly classified soil pH in 97% of cases while reducing the analysis turnaround time from days to minutes. Moreover, by performing on-the-spot analyses of individual compound sub-samples in the field, we have achieved a 9-fold increase of spatial resolution that reveals pH-variations not detectable in compound mapping mode. Our mobile system can be extended to perform multi-parameter chemical tests of soil nutrients for applications in environmental monitoring at marginal manufacturing cost.**

Chemical analysis of soil and water is essential for informing agricultural decision-making[1-3]. In precision agriculture, agronomists routinely collect soil samples which are transferred to specialized labs with dedicated equipment operated by trained experts. The cost and time for performing laboratory analysis, however, limits application in emerging economies where lab infrastructure and equipment are sparse and the costs prohibitive. Sample transfer to the lab is time consuming, expensive and, in the case of cross-border shipments, regulatorily complex. Portable analysis kit or electronic sensors used for on-site chemical monitoring could provide a much-needed alternative. However, they are impractical or unaffordable for smallholder farmers with an annual production equivalent of a few hundred US dollars or less as is often the case in agriculture-based economies. As a result, those farmers may be cut off from the benefits of up-to-date chemical data needed to improve their agricultural production program.

High-resolution chemical data made available through lab analysis is, however, not necessarily needed to detect existing nutrient deficiencies or abnormal acidity in soil that require immediate remediation action. Even a binary chemical test result of the *below-above* type, if referenced to a suitable threshold, can be valuable for informing decision-making processes in the field. As an example, if optimum soil pH is defined by, say, *pH above 6*, a test result stating - *pH below 6* - could inform corrective soil treatment until optimum soil conditions - *pH above 6* - are restored. With a mobile test system that measures against such threshold levels at marginal cost to the farmer, soil health would become "micro-manageable" in near real-time. This would improve agricultural efficiency while, at the same time, excessive use of chemical product and the related environmental damage could be avoided. Besides the time advantage in obtaining chemical results in the field, soil sample shipments could be limited to cases where high-resolution data and in-depth agronomical analysis are required.

From a technological point of view, recent advances in mobile communication systems have made possible the integration of paper-based sensors within a high-tech/low-tech hybrid approach[4-14]. A smartphone, the high-tech device, can be configured to perform the readout and analysis of a colorimetric paper sensor, the low-tech device, without added hardware features and at virtually no cost to the user. For moving hybrid soil testing technology from lab to field, we have just recently removed two major roadblocks. Firstly, demonstrating AI-based colorimetric calibration models that adapt to the edge computing capabilities of standard mobile phones[15] and, secondly, resolving the issue of ambient light correction[16], a prerequisite for mobile paper sensor readout under challenging light conditions in the field. For managing field locations, the integration of GPS services enables spatial resolution down to the level of meters at any given location while cloud integration enables field data analysis and visualization at scale. Based on these developments, it is conceivable that a mobile phone user could now perform a soil test with a paper sensor at a pre-defined GPS location, retrieve the chemical results immediately to perform remediation action locally and, finally, stream the measured test data to a cloud database for spatiotemporal analysis. In the following, we demonstrate the first application of the above workflow to soil pH assessment at field scale.

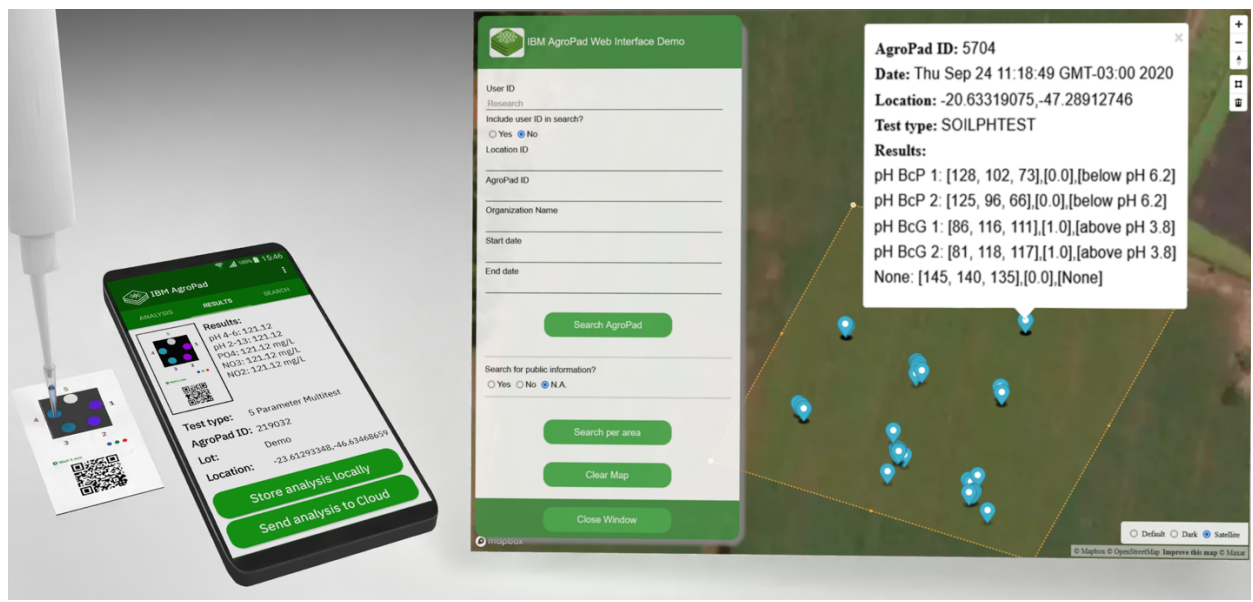

**Figure 1. Mobile chemical analysis system.** Artistic rendering of the mobile soil measurement system. A liquid sample of soil extract is deposited onto a paper sensor. The paper sensor contains colorimetric indicators for producing color output as function of chemical concentration in the soil sample. The sample-specific color output is then analyzed by a mobile phone application and, by means of integrated AI-based models, converted into chemical concentration. The chemical information is merged with test information, location information, and time information for local storage or network transfer. **Inset** Web interface for visualizing soil pH test results based on GPS location data which is available at https://agropad-demo.mybluemix.net/. The data are retrieved from a cloud database which integrates the individual measurements performed with a mobile phone in the field.

In Figure 1, we show the measurement system developed for this field study. The main components are a paper sensor with integrated colorimetric indicators that react upon deposition of a sample of liquid soil extract at the front side and provide a concentration-specific color in the output area at the back side. A research prototype software application deployed in a commercially available smartphone acquires an image of the output layer of the sensor through the mobile phone's camera. The app software performs a sequence of analysis steps, i.e., image segmentation, color extraction, and application of colorimetric calibration models, to process chemical concentration results. Within an instant, the soil analysis results are available to the app user and, tagged with time and location information, stored locally, or transferred to a cloud computing platform for data integration, analysis, and visualization. The measured data can be visualized through a web-based user interface. We briefly outline in the following the advances made in our research and development of next-generation paper sensors, and AI-based colorimetric data analysis and how their combined application has enabled a first successful benchmark study at field scale.

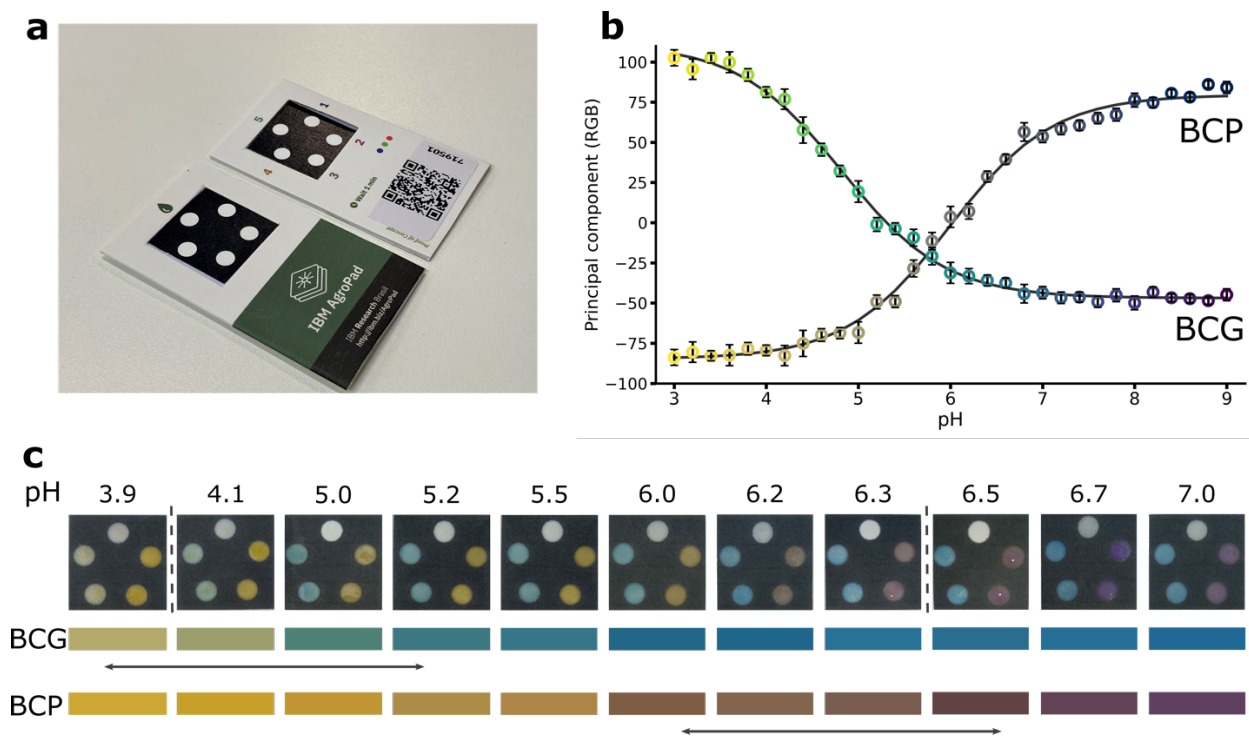

**Figure 2. Paper based, colorimetric soil sensor device. a** Research prototype of the paper-based chemical sensor used in this study with sample deposition layer (front) and color output layer, reference marks and QR code (back). **b** Principal components of measured color output (symbols: experimental data, lines: sigmoid fit functions) obtained for colorimetric indicators Bromocresol Green (BCG) and Bromocresol Purple (BCP), respectively, as function of sample pH across the entire range of the paper sensor. The dashed lines separate classification results with regards to low pH (pH 3-3.9), medium pH (pH 4.0-6.3), and high pH (pH 6.4-9). **c** Representative photos of the output layer area of the paper sensor depicted in **a**. Depending on the pH value of the deposited soil sample, the output layer features the characteristic color of integrated colorimetric indicators (circles 1, 2: BCP; circles 3, 4: BCG, circle 5: reference w/o indicator). The bar charts visualize the colorimetric outputs obtained in the different pH regimes.

In Figure 2, we show the paper sensor used in this study. Vertically integrated, microfluidic paper analytical devices (μ-PAD) devices[17-21] have been successfully demonstrated for water and soil analysis under laboratory condition[8,22-27]. For field application, we have developed a novel multi-indicator µ-PAD with soil filter function, embedded color reference features for ambient light correction, and QR-encoded data for automated processing through the mobile app. The sensor integrates two colorimetric pH indicators, Bromocresol Green (BCG) and Bromocresol Purple (BCP), for creating pH-specific color output. We have developed artificial intelligence/machine learning (AI/ML) models that transform the color produced by the two indicators as function of pH into soil pH results while correcting for the challenging ambient light conditions in the field. The accessible measurement range of the system is pH 3-9, with three classes – low pH (pH 3-3.9), medium pH (pH 4.0-6.3), and high pH (pH 6.4-9). We provide technical details with regards to the paper-based sensor device and the AI/ML calibration models in the Methods Section and the Supplementary Information.

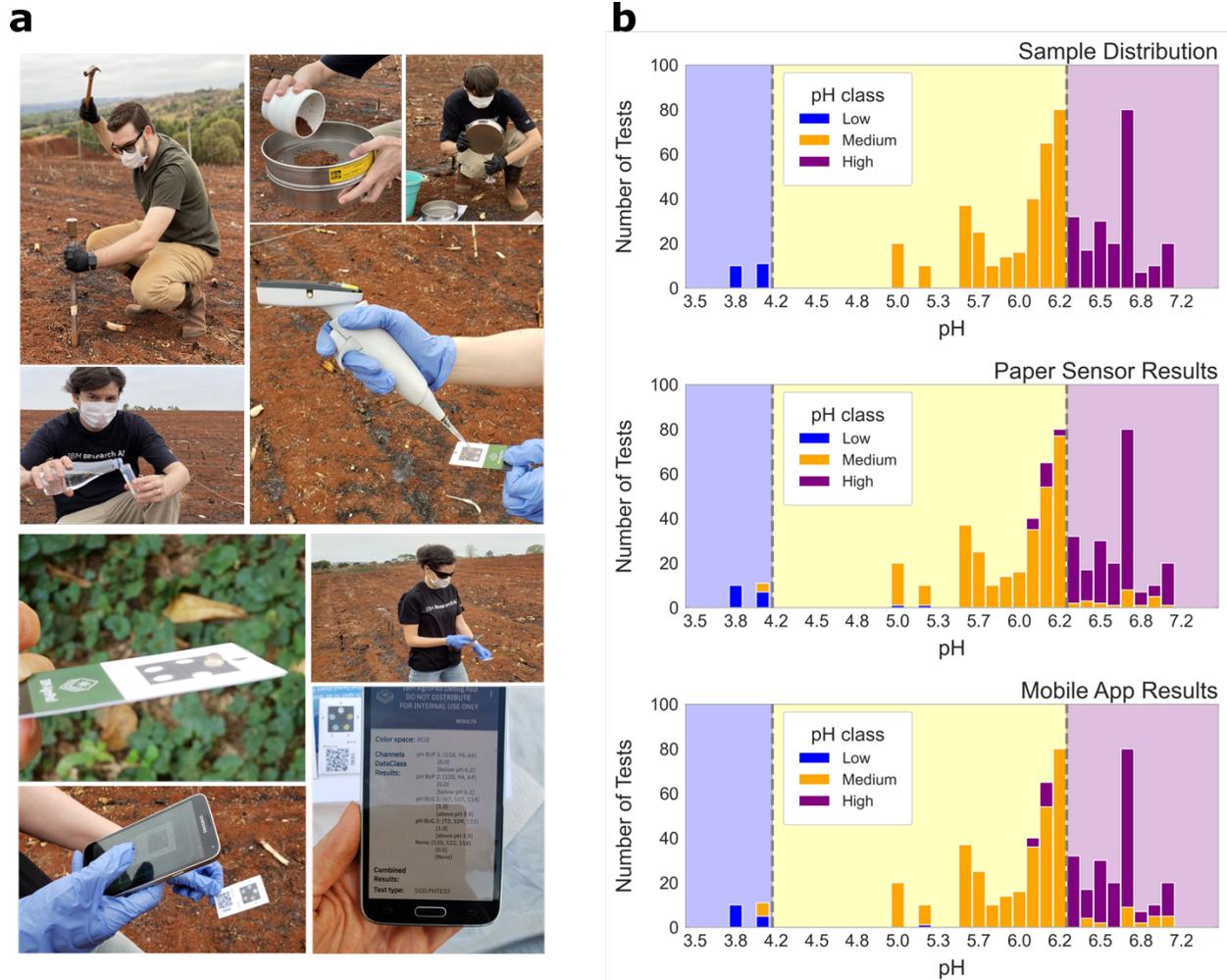

**Figure 3. Mobile soil pH testing in the field. a** Soil sample collection, processing, and mobile phone assisted readout of the paper sensor in the field. **b** Soil pH field data. Upper panel: Number of soil pH occurrences for all samples included in this study (N=548). Medium panel: pH classification results as per visual analysis of the paper sensor test card output. Lower panel: pH classification results as per automated, mobile app assisted analysis of the paper sensor test card. The dashed lines separate classification results with regards to low pH (pH 3-3.9), medium pH (pH 4.0-6.3), and high pH (pH 6.4-9).

In Figure 3, we show the field application of the mobile measurement system for soil analysis. We have collected and processed soil samples, including a soil pH extraction step, at pre-defined locations in the field (see Methods Section and Supplementary Information). Deposition of a small volume of soil extract on the paper sensor is followed by a wait time of about 2 mins before the mobile app is applied to perform the test. For all soil samples studied, we have independently determined their pH values with a complementary standard reference analysis technique. Overall, we have performed a grand total of 805 paper sensor soil tests to confirm the accessible measurement range and accuracy of the mobile measurement system and we have successfully detected instances within all three pH classes. This includes 615 tests performed in the field and 190 tests performed under laboratory conditions. For separating the error analysis of sensor device and mobile app, we have performed a visual analysis of each sensor to verify if the color output represents the actual pH value of the sample. We have then applied the mobile test system to the same sensor devices for analyzing the readout accuracy of the mobile app. As a result, based on the visual inspection, we obtain an overall paper sensor test accuracy of 73% (590 Out of 805). This means that, on average, about three out of four paper sensors have developed a proper colorimetric reaction and produced a valid test result. The overall, mobile phone assisted readout accuracy of the paper sensor test cards is 72% (579 Out of 805). If we perform the mobile phone assisted readout with a subset of paper sensors that have produced a proper colorimetric output as confirmed by visual inspection, we obtain a mobile phone assisted readout accuracy of 92% (505 Out of 548). On average, about nine out of ten properly functioning paper sensor test cards are read out and classified correctly by the mobile app, attesting to the app's reliability of both colorimetric calibration and ambient light correction. We conclude, therefore, that the mobile test systems' overall accuracy is currently limited by the paper sensor performance.

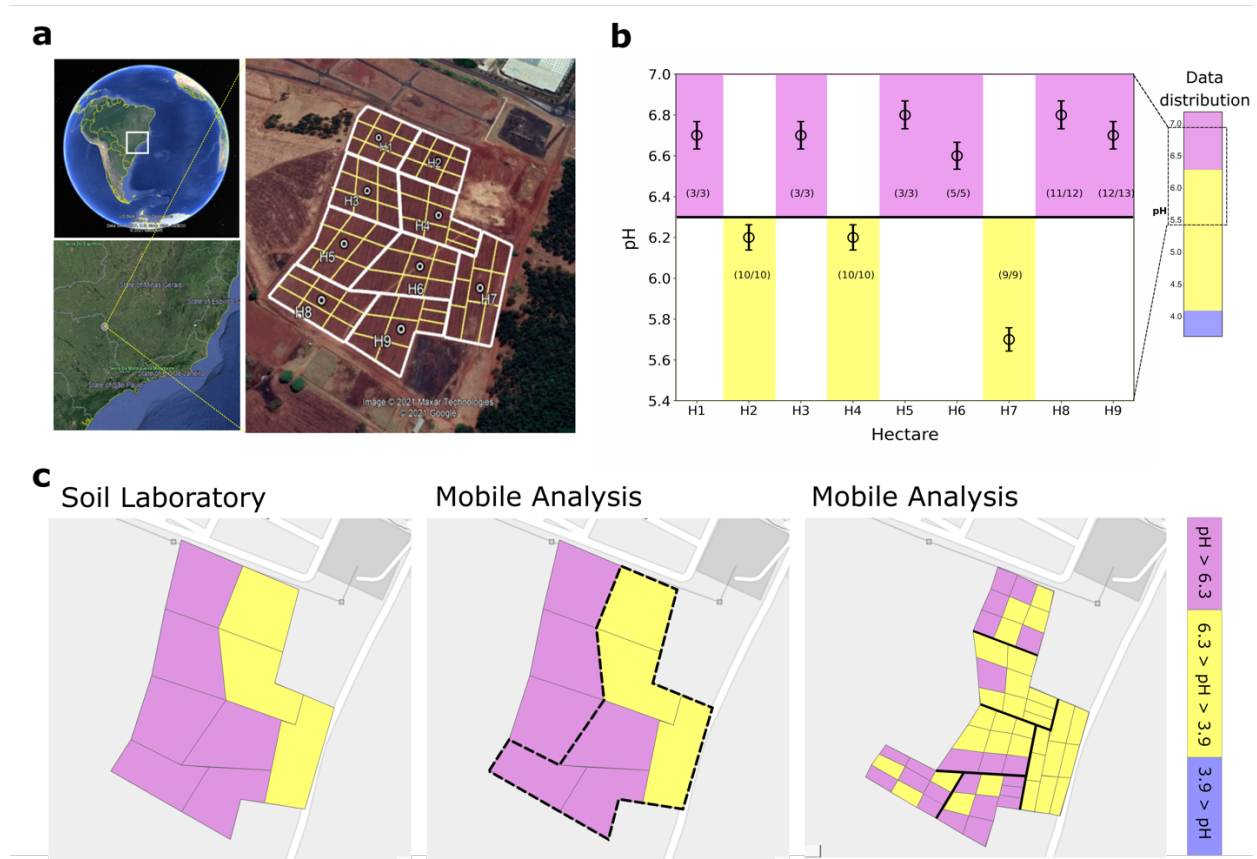

**Figure 4. Soil pH benchmark field study. a** Location of the testing site and demarcation of soil sample collection zones labeled H1-H9. **b** Soil pH results obtained with mobile analysis of paper sensors versus values obtained from routine soil analysis in the lab for the same compound samples collected in H1-H9. The graphic on the right-hand side highlights the relevant pH-range, 5.4-7, and the pH threshold at 6.3 which separates the classes "medium soil pH" and "high soil pH". The symbols represent the pH of the compound samples as measured by the soil analysis lab, and the number in parenthesis represent the ratio between correct mobile test results and total mobile tests performed for a specific compound sample. **c** Spatial map of soil pH visualizing the compound sample pH classification derived from routine lab analysis (left) and pH classification of the same samples obtained with the mobile test system (middle). The bold dashed lines (middle) delineate the area for which soil samples were analyzed with 9x higher spatial resolution (right) as compared to standard protocol. The paper sensor results (right) reveal the spatial fine structure of soil pH distribution at the test site.

For benchmarking the paper sensor test system against standard lab analysis, we have identified a suitable agricultural operation in the state of São Paulo, Brazil. In Figure 4, we show the location and delineation of the test site with a total area of 9 hectares. For the collection of compound samples and the shipment of samples to the lab, we have followed precision agriculture protocol: Each cell having a size of about 1 hectare is subdivided into 9 sub-zones. In each of the nine sub-zones, distinct locations are marked via GPS signal for sample collection. The compound samples representing each of the nine cells, each of which are composed of samples of their respective nine sub-zones, are sent to a soil analysis lab. In parallel, we analyze the same compound samples with our mobile pH analysis systems. For comparison, the pH results retrieved from the soil analysis lab are classified with regards to the pH thresholds of the mobile test system.

As a key result of our study, we find that our mobile test system result correctly predicts the pH class obtained with the soil lab results for all nine cells investigated, resulting in 6 cells with "high pH", 3 cells with "medium pH", and zero cells with "low pH". Through repeat measurements, we find that the mobile system differentiates correctly between "medium soil pH" and "high soil pH" in 66 out of 68 cases, leading to an overall measurement accuracy of 97%. We would like to point out that we have independently validated the pH values obtained with the field-based soil extraction method. As shown in Supplementary Figure S7b, we have measured the pH of all soil extracts using a standard reference method and we have observed quantitative agreement with the pH results obtained from the soil analysis lab.

For 6 of the 9 cells studied, we have performed mobile pH analysis at each of the GPS locations, directly in the field. The results provide a mapping of soil pH at 9x higher spatial resolution as compared to the compound sample analysis performed at cell level. Indeed, the chemical mapping at higher spatial resolution potentially improves soil health management efficiency: If pH corrections were to be performed based on the 9x resolution mapping, an additional 16 out of the total 54 sub-zones, or 30% of the area, would be treated accurately as compared with the mapping in standard resolution. By assuming soil treatment would be performed to increase pH from "medium pH" to "high pH", about 22% of soil correction product could be saved for a total of 6 subzones that would not require soil pH correction in this scenario. The resolution advantage could particularly benefit farmers with smaller fields and limited resources for soil correction.

Finally, we analyze potential time and cost advantages realizable through paper-based analysis. The time to perform the mobile analysis in the field is of the order of 20-30 mins per sample, mainly due to soil extract preparation, so tests can be performed at each GPS location or batch processed at a central field location if GPS coordinates are recorded for each sample. In our study, this has led to a mobile test turnaround time of less than 1 hour per cell (after soil extraction), as compared to the standard lab test turnaround time which is on the order of days or weeks, depending on the infrastructure. We estimate paper sensor manufacturing costs of about US$3.00 per piece at the current R&D scale (thousands of devices) and below US$0.25 per piece at larger scale (millions of devices), potentially enabling an inexpensive alternative to high-end lab analysis. As shown in Supplementary Figure S2 and Table S1, we have performed manufacturing scaling studies that resolve existing limitations in the paper sensor production workflow. It is important to note that the paper sensor platform can integrate a broad range of colorimetric indicators for adaption to specific use cases[5,7,8,12,14,20,22,23,27]. As a multi-parameter sensor example adapted to agricultural requirements in Southeast Brazil, we show in Supplementary Figure S12 a paper sensor prototype that, in addition to pH, integrates colorimetric indicators for the detection of Magnesium, Calcium, and Aluminum ions. Future research work requires the development of colorimetric calibrations and soil extraction methods to support field-scale benchmarking of a broader class of soil nutrients and chemical parameters for enabling adoption of the technology in emerging economies.

In summary, we have reported the first mobile chemical analysis of soil with paper sensors at field scale. The method produces reliable soil pH results in near real-time with potential applications in agriculture and environmental monitoring. The mobile soil analysis system and the paper-based device have the potential to integrate further test parameters such as, for example, Magnesium, Calcium and Aluminum, which are currently being evaluated and tested. If produced at mass scale, the paper-based chemical sensor could become a technologically viable, low-cost alternative to lab testing. Considering a rising adoption of smartphones globally, paper-based test applications serving smallholder farmers in emerging economies could soon become a reality.

## MATERIALS AND METHODS

### Paper-based test device design and manufacturing

For field testing, we developed a prototype device layout consisting of two identical paper layers (3001-861 CHR1 200x200 Chromatography Paper, Whatman) with 5 isolated hydrophilic spots per layer. As shown in Supplementary Figure S1, one layer acted as sample input and filter layer and the other one as reaction and output layer. The microfluidic patterns were printed onto the paper surface by a wax printer (ColorQube8580, Xerox) and heated to 100°C for 1 minute to create the hydrophobic barriers that confine the liquid in the desired flow path. Once the barriers were defined, we pipetted 2 μL of colorimetric indicators (Bromocresol Green and Bromocresol Purple, Quimlab) onto the circular chambers of the output layer. After a drying step, we aligned and glued the sheets together (77 Spray Glue, 3M). The paper-based, microfluidic devices were then packaged within a cardboard cover to increase test robustness and for adding the QR-code identifier, image processing markers, and labeling.

Once manufactured, we bundled the devices up in batches of 50 and vacuum-sealed them within transparent plastic bags to increase shelf life. To ensure proper performance, we have monitored the colorimetric output of devices over a period of 4 months. To avoid potential device deterioration, we fabricated the paper sensors used in our field study within a week prior to usage, vacuum-sealed and transported to the test site without noticeable degradation. During field testing, devices were exposed to the varying weather conditions with a maximum temperature of 31º C. We note that varying temperatures could impact the amount of sample needed for producing a colorimetric response in the device (due to variations in sample evaporation time). Further information regarding lab-scale and large-scale device production is provided in the Supplementary Information, including Supplementary Figure S2 and Supplementary Table S1.

**Mobile phone and cloud computing applications**

To enable automatized collection, analysis, and storage of soil pH test result, we processed the colorimetric test output by a mobile phone application and transferred the data to a cloud computing application for data integration, analysis, and visualization, see Supplementary Figure S3. To enable automatized readout with the mobile app, each device carried a QR-code sticker.

The QR-code contained a unique test identifier along with information regarding the manufacturing lot, the paper device configuration, and the chemical indicators.

In our field study, we used a mobile smartphone (Galaxy S5, Samsung) for test readout. We processed the acquired images following the workflow depicted Supplementary Figure S3 with a dedicated mobile application (operating in Android, Google). The test results, together with the geolocation, timestamp, the unique ID, and raw images of the test card, were immediately available to the user through the user interface of the smartphone application. In addition, the data were either saved locally on the phone or sent via network connection to a no-SQL database in the cloud (Cloudant, IBM) via an API Connect implementation. Supplementary Figure S3c displays an example json-file representing a single soil measurement. It contains the soil pH class result as processed by the mobile application, the colorimetric information, the geolocation and timestamp, and as-acquired raw image data for reference.

**Mobile app-based data acquisition and processing workflow**

The data acquisition and processing workflow is schematically visualized in Supplementary Figure S4. The mobile soil test and data acquisition process is illustrated in Supplementary Figure S4a. The mobile application acquired a single image of the sensor output and the QR-code. The QR code contained the test ID and ensured that the appropriate AI/ML models and configurations were used for data processing. In a first step, as indicated as indicated in Supplementary Figure S4b, the image processing routine located and segmented both the microfluidic device area where the sensor output spots were located and the color reference area with color spots were used as input for compensating ambient light conditions. Subsequently, color channel information (RGB) from each individual circle on the paper analysis card as well as those of the color reference scale were extracted.

We developed algorithms for compensating variations in data acquisition conditions including device positioning, image contrast and sensor exposure so to ensure reliable segmentation outcomes under field conditions. To ensure that results were processed regardless of acquisition hardware or illumination conditions, we developed a light compensation algorithm[16]. The algorithm used the color references on the test card to calibrate the mean RGB value of each colorimetric output, enabling mobile pH testing even under challenging light conditions. Finally, as shown in Supplementary Figure S4e, the corrected RGB values were transformed into pH classes by a set of AI/ML models.

**Colorimetric indicator selection and creation of AI/ML calibration models**

To develop a paper device for analyzing soil samples over a broad pH range of 3-9, we evaluated various standard pH indicators including bromocresol green, bromocresol purple, bogen and methyl red. Specifically, we tested the response of the indicators within the paper substrate by manufacturing and treating devices with different concentrations and volumes of the above chemical indicators. For each device, we treated all reaction spots with the same indicator concentration, and, after a drying step, we applied 5 to 10μL of a standard buffer solution to each spot. Once the solution had reacted with the buffer for several minutes, we measured their colorimetric response under controlled light, temperature, and humidity conditions with the image acquisition system displayed in Supplementary Figure S5a. The system consisted of a machine vision camera (PL-D734CU-T, Pixelink) equipped with a telecentric lens (Newport) and homogeneous illumination source (LED144A, AmScope). Also shown in Supplementary Figure S5a is a calibration paper device, having two spots with BCG indicator and two spots with BCP indicator, respectively, mounted in a sample holder containing the color references needed for model calibration.

For training and testing AI/ML models that transform RGB values into pH classes, we acquired several datasets of images for each measurement condition, i.e., choice of colorimetric indicator, indicator concentration, and volume, reaction time, as a function of the pH-value of the buffer solution. After collecting the images, we extracted the mean RGB value per color spot. Subsequently, we applied principal component analysis (PCA) to reduce dimensionality and to evaluate the colorimetric response as a function of pH, as shown in Figure 2b of the main manuscript.

For our field test application, we defined three pH classes for analysis: a "low pH" class for values below 3.9, a "high pH" class for values above 6.3, and a "medium pH" class for the values in between. For best classification results, we chose bromocresol green (BCG) with color change occurring at pH=3.9, and bromocresol purple (BCP) with a color change occurring at pH=6.3, as shown in Figure 2 of the main manuscript. We set the sample volume and reaction time to 5 mL and 2 minutes, respectively, to guarantee that the colorimetric reaction was completed and to minimize reflection-induced variability in the image data caused by excess liquid.

We collected colorimetric data with a set of 14 soil extract samples having pH-values ranging from 3.4 to 6.9 which were prepared by following the field extraction protocol provided in the Supplementary Information. Supplementary Figure S5b displays a representative image data set. We extracted mean RGB values for each color spot. Supplementary Figure S5c shows the distribution of the calibration data in the RGB space and the way the data were divided in two classes per indicator. Apparent color changes occured at pH 3.9 for the BCG indicator and at pH 6.3 for the BCP indicator, respectively. The combined analysis of two sensor spots per indicator increased test robustness. We derived the final pH test result from the individual classification results of the BCG and BCP indicators by following the logic laid out in Supplementary Figure S5d.

We used OpenCV (https://opencv.org/) which operates well within the computational limitations of edge devices and supports offline operation at locations that lack network connectivity. We compared test image processing results obtained with script implementations in Java and Python, respectively, and obtained agreement. Within a data cleaning and curation step, we split the data set of each reagent into two classes. For BCG, we obtained 160 data points for class 0 and 430 data points for class 1. For BCP, we obtained about 460 data points for class 0 and 100 data points for class 1. For model training, we used as features either a RGB vector per data point, its transformed HSV vector, or a combination of them. 80% of the data were used for training while 20% were used for testing. We repeated training five times, splitting the dataset randomly by 80:20, and averaged the accuracy score in 5-way cross-validation. We trained logistic regression models as well as support vector machine models, optimizing the hyperparameters to reach a test classification accuracy close to 100%. Prior to field application, we confirmed the accuracy of the best-performing classifiers on colorimetric data acquired with 15 test devices using the mobile phone application at field conditions, achieving consistent accuracy better than 85%.

We merged the parameters of select models for various indicators into json-files, along with information about the chemical indicators, the classification labels, device lot, and light compensation references. As depicted in Supplementary Figure S5e, the files were stored in a cloud database from which the mobile phone application retrieved and applied the appropriate model during operation. This workflow allowed for model updates without changing the mobile phone application.

**Soil sample collection, soil extraction, lab benchmark, data analysis**

We documented the site conditions, soil sample collection and field test execution, including soil extraction protocol and soil lab benchmark description, in dedicated subsections of the Supplementary Information, including Supplementary Figures S6 and S7. Also, the in-depth analysis of paper-based pH measurements is described in a dedicated subsection with Supplementary Figures S8-S11.

**Multi-parameter soil test**

A paper-based sensor prototype integrating colorimetric indicators for the simultaneous detection of soil pH, Aluminum, Calcium and Magnesium ions is shown in Supplementary Figure S12.

**DATA AVAILABILITY**

The field test data acquired in this study and reported in the manuscript is available (under DOI: 10.24435/materialscloud:vt-4t) at: https://archive.materialscloud.org/record/2022.91

The test data can be visualized with a data mapping interface which is implemented based on web-services (Maps API, Google) at: https://agropad-demo.mybluemix.net/

The web application retrieves the data based on the test identifiers provided in the spreadsheet column labeled "upadID". When clicking on the marker of a given measurement, a pop-up window provides the test result along with measurement time, location, and colorimetric information.

**CODE AVAILABILITY**

The algorithms for processing the field test data acquired with the paper-based sensing devices, including image feature extraction and colorimetric analysis, are available as Python code at:

https://github.com/IBM/paper-device-colorimetric-analysis

The code repository contains Jupyter Notebooks for simplifying data processing and visualization. Test data in suitable input format are provided through the link in the Data Availability section.

**ACKNOWLEDGEMENT**

We thank Emanuel Carrilho and Cleyton Nascimento Makara (University of São Paulo - São Carlos) for discussions of paper based chemical sensors, Marcelo Ramos and Fabio Correa (3RLAB/REHAGRO) as well as Daniel Klein and Josilaine Prado (TIMAC) for discussions on soil analysis and the provision of soil samples. We acknowledge significant contributions to the project by Jessica Barbosa Lima (formerly at IBM), Mariana Brito Rodrigues and Paula Fernanda Pereira (both IBM). We further acknowledge project support by David Browning (ENVERITAS),

Tiago Alves (CSEM BRASIL), Vossie Wilsnach (OMNIA FERTILIZERS RSA), Adrienne Sabilia, Chris Sciacca (both formerly at IBM), Juliana Setembro, Alexandre Pfeifer, Raquel Chebabi, Ulisses Mello and Bruno Flach (all IBM).

# Artificial Intelligence enables mobile soil analysis for sustainable agriculture

# -SUPPLEMENTARY INFORMATION-

## Paper-based test device design and manufacturing

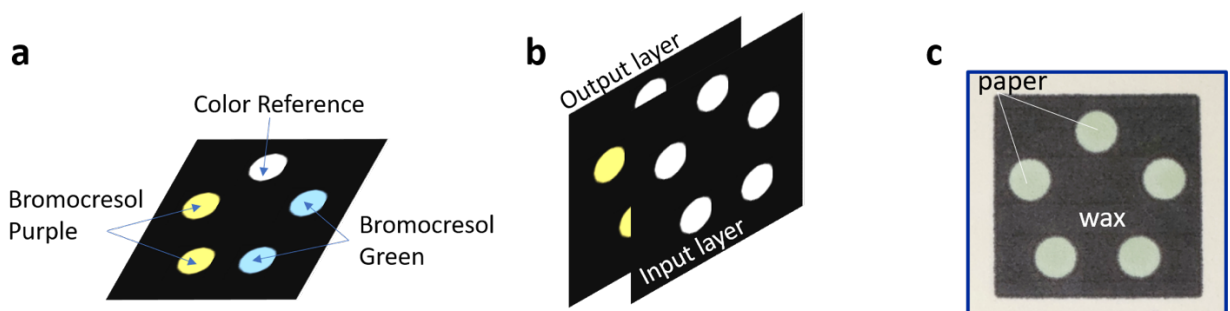

**Figure S1.** (a) Layout with indicator distribution of the paper-based soil analysis device. (b) Assembly of device layers. (c) Photograph of a device prototype.

The test sensors consisted of a two-layer, microfluidic paper-based analysis device (µPAD) shown in Figure S1 enveloped within a carboard cover with graphic features to support test readout. Figure S2a visualizes the key device production steps of the test devices at laboratory scale. In a first step, we printed the test layout of the microfluidics device onto the surface of chromatography paper with a wax printer. In a reflow step, we heated the paper in a hot press to melt and define the hydrophobic barriers that control the liquid flow. We then pipetted the colorimetric indicators on the wax-defined hydrophilic areas. After a drying step, we vertically aligned and assembled the paper layers using stainless-steel masks to protect the hydrophilic areas. Finally, we sandwiched the paper-based test device between top and bottom covers fabricated in water-resistant cardboard material. In a final step, we attached a QR code sticker to the device backside where colorimetric detection and mobile readout is performed. This method provided a reliable and reproducible manufacturing process for making thousands of test devices in our lab.

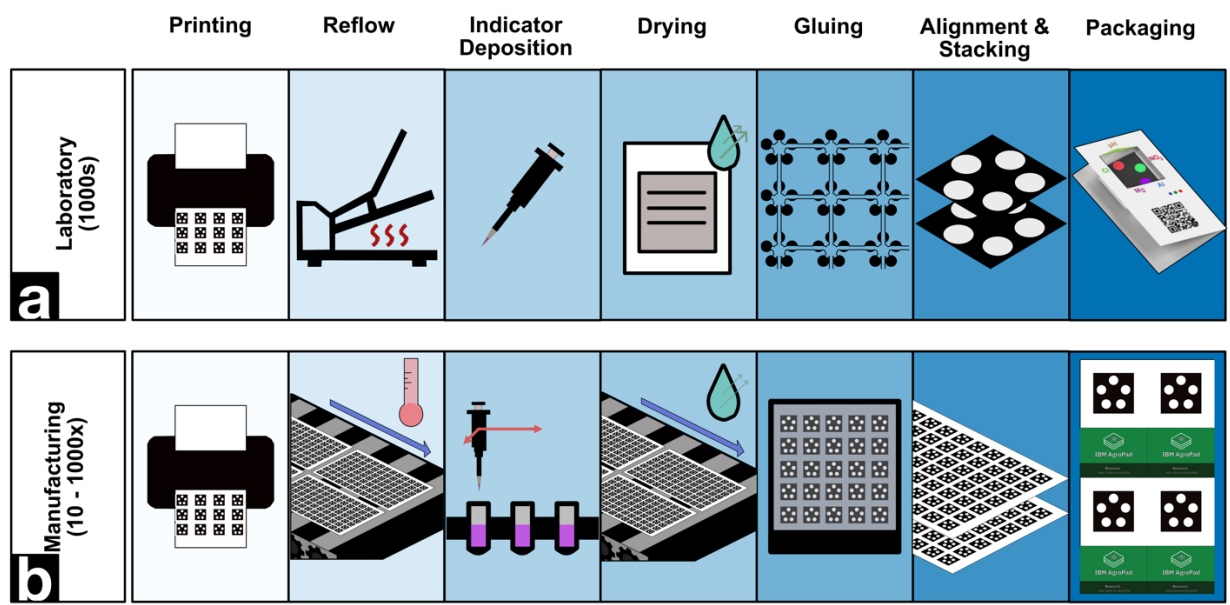

**Figure S2.** Production steps of microfluidic paper-based devices at (a) laboratory scale and (b) industrial scale.

To explore the scaling potential of paper sensor manufacturing, we have performed a production study. Figure S2b illustrates the main steps of the scaled production workflow. While the wax printing step remains essentially unchanged, the main manufacturing bottlenecks occur in the deposition of the chemical indicators and in the device assembly. In our production study, we consider the same class of wax printers used in lab-scale manufacturing to print the test layout on paper. We note that parallel operation of multiple printers is possible.

Subsequently, the wax-printed sheets are heated as they travel through an in-line reflow oven. For deposition of liquid indicators, we insert the paper sheets into automated reagent deposition equipment. The sheet drying step is carried out in another in-line oven set to a lower operating temperature. We deposit glue by means of serigraph printing, where liquid glue is pressed through a stainless-steel mask to avoid exposure of the reaction zones. The sheet alignment and stacking steps are carried out with the help of fiducial markers or holes in the paper sheets. Once aligned and sandwiched, the sheets are cold pressed to ensure the contact between the layers is homogeneous. Finally, individual µPADs are cut out at sheet level with a cut and crease machine.

We have evaluated all the above steps on the production floor, except for automated pipetting of indicators. The cardboard covers are manufactured following standard scaled printing and cutting process while QR code stickers are added at sheet scale. Individual µPADs are then aligned between sheets of front and back cardboard covers, glued, and pressed in place before vacuum sealing the devices either individually or in bundles for distribution.

The production program we have explored for a 2-layer paper-based device with an area of 24x24mm$^2$ is detailed in Table S1. Assuming an 8-hour production period within an industrial manufacturing environment, this would enable monthly production of +300.000 devices, considering one production period/day, one machine, 22 working days and using the bottleneck process as a reference.

| Item | Category | Process Description | Machine Description | Estimated Production Rate (Per min) | Estimated Sensor Production Per hr | Estimated Sensor Production Per 8hrs |
|---|---|---|---|---|---|---|
| 1 | Paper Sensor | Printing | Wax Printer | 10 sheets | 10800 | 86400 |
| 2 | | Channel formation | Reflow Oven | 1 meter | 5400 | 43200 |
| 3 | | Reagent deposition | Automated Liquid Handling | 300 depositions | 2160 | 17280 |
| 4 | | Sheet drying | Reflow Oven | 1 meter | 5400 | 43200 |
| 5 | | Gluing | Screen Printer | 5 sheets | 10800 | 86400 |
| 6 | | Alignment & Stacking | Customized Equipment | 3 sets | 6480 | 51840 |
| 7 | | Lamination | Lamination | 10 sheets | 21600 | 172800 |
| 8 | | Cutting | Cut & Crease Machine | 6 sheets | 12960 | 103680 |
| 9 | Cover | Printing | Digital Press | 120 sheets | 115200 | 921600 |
| 10 | | Cutting | Cut & Crease Machine | 6 sheets | 11520 | 92160 |
| 11 | | Gluing | Glue Machine | 20 meters | 89600 | 716800 |
| 12 | Assembly | Chip alignment & cover stacking | Customized Equipment | 1 set | 1920 | 15360 |
| 13 | | QR Code | Label Printer | 6 meters | 43200 | 345600 |
| 14 | | Packaging | Vacuum Sealer | 4 bags | 2400 | 19200 |

**Table S1.** Study of scaled manufacturing of paper sensor test devices.

## Mobile phone and cloud computing applications

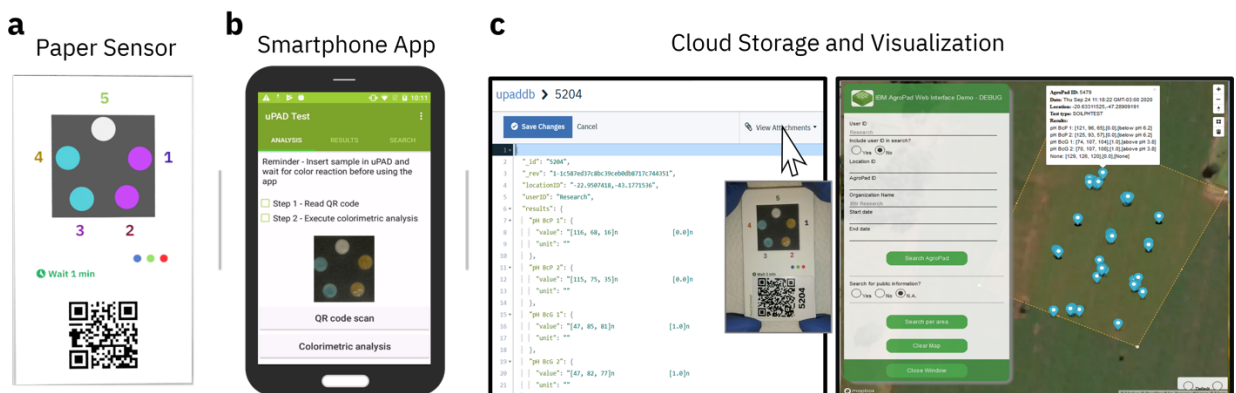

**Figure S3.** Mobile chemical analysis system components. (a) Sensor test card comprising a microfluidic paper-based analytical device, a QR-code and color correction references. (b) Screenshot of the mobile phone application for test data acquisition, processing, and upload. (c) Screenshot of (left) a soil test entry in the cloud computing data base and (right) and visualization interface with soil data overlaying the map of a test site.

## Mobile app-based data acquisition and processing workflow

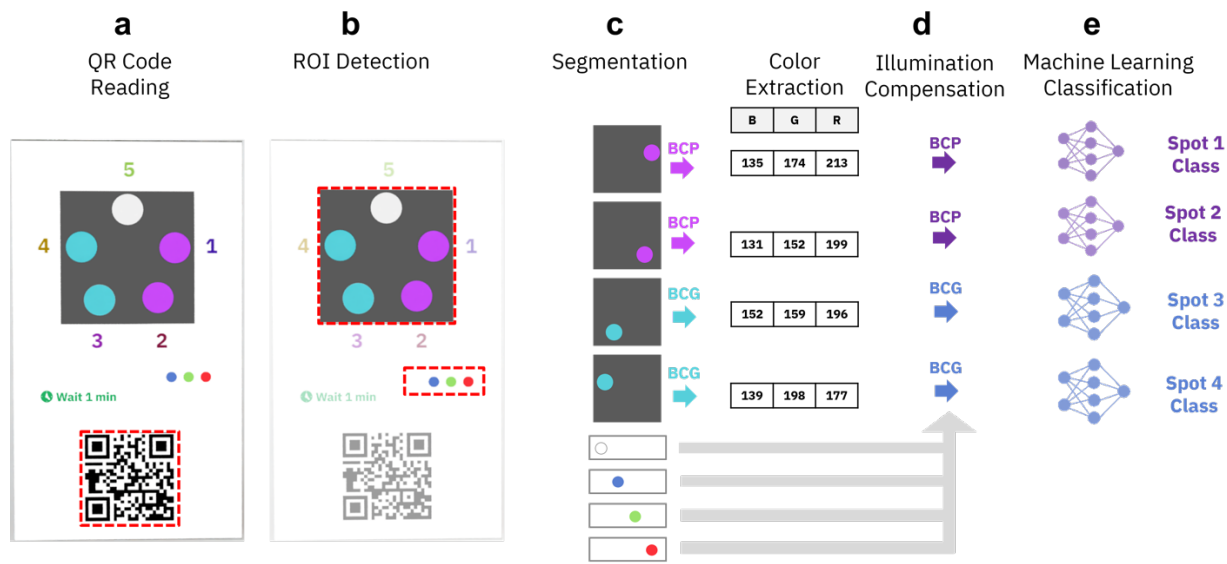

**Figure S4.** Mobile application-based data acquisition and processing workflow. (a) Image capture with QR-code readout which is highlighted by dashed red lines. (b) Region-of-Interest (ROI) contains sensor output and color correction references as highlighted by dashed red lines. (c) Segmentation of colorimetric sensor output spots and color scale references and computation of the mean RGB values for each spot. (d) Illumination compensation steps. (e) Machine learning classification step to the extracted color data for producing pH results.

## Colorimetric indicator selection and creation of AI/ML calibration models

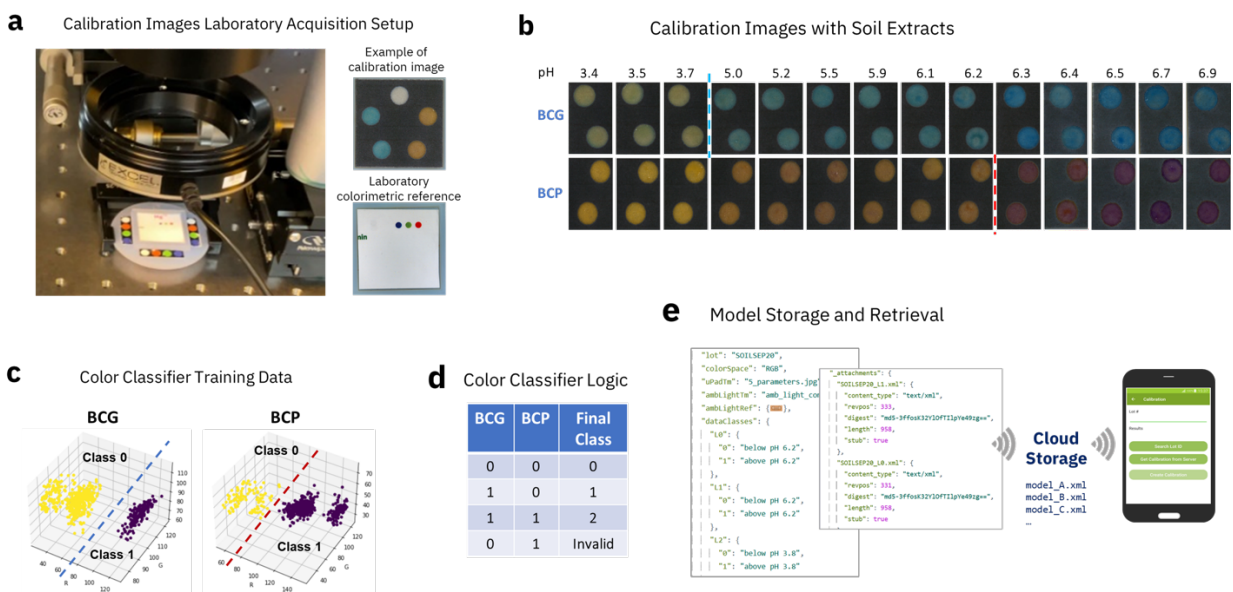

**Figure S5.** Colorimetric calibration setup (a) Experimental setup for acquiring colorimetric calibration data. (b) Measured soil extract calibration images at various pH levels. (c) Distribution of the measured data in RGB space. (d) Logic for combining the classification result of the BCG and BCP indicators. (e) Cloud database for storage and data retrieval by the mobile phone app.

# Soil sample collection, extraction, lab benchmark, data analysis

## Soil Sample Extraction Protocol

The analysis of soil through colorimetric reactions requires extracting soil nutrients from soil samples in liquid phase by means of chemical reagents. In the application of paper-based devices, the papers' cellulose fibers need to withstand the chemistry used for nutrient extraction.

Various extraction methods exist, each with parameters that can influence the colorimetric response of the reaction: i) the soil chemical property to be analyzed, ii) the type, volume, and concentration of the reference solution; iii) the volume of the soil sample; iv) the procedure time; v) the tools to be used, and vi) the soil buffering capacity [Motsara08, Hazelton16, Teixeira17]. Buffering capacity refers to the ability of soil to resist changes in pH and increases with cation exchange capacity and organic matter content [Hazelton16]. This parameter requires particular attention so to ensure consistent relationship between nutrient concentration and colorimetric result. In practice, depending on the soil's pH and its buffering capacity, the pH stabilization of the sample after extraction, i.e., soil extract formation, can take several minutes. Significant pH variations typically occur within 15 minutes after the soil sample is mixed with the extractor.

For our field study, we followed the extraction protocol for measurement of soil pH issued by EMBRAPA [Teixeira17]. The protocol consists of sieving the sample through a 2mm mesh sieve, mixing the result with 0.01M $CaCl_2$ solution in a 1:2.5 ratio in a vial with a cap, shaking it for 60 seconds and letting it settle for at least 15 minutes. To test and validate the extraction protocol for application within the paper device, we used a variety of reference soil samples from various locations in Brazil. The samples were collected from topsoil, down to 20 cm below the surface, dried in natural air and sieved following the same protocol for consistency. After mixing the samples with a $CaCl_2$ solution of pH=5.5, pH measurements were taken with a pH-meter (Simpla 140, AKSO) at distinct time intervals. In Figure S6a, we plot the average pH value obtained from 10 soil pH extraction procedures, with the same soil sample, together with the respective standard deviations. Based on the results, we have chosen the minimum extraction time in the field to be 20 minutes.

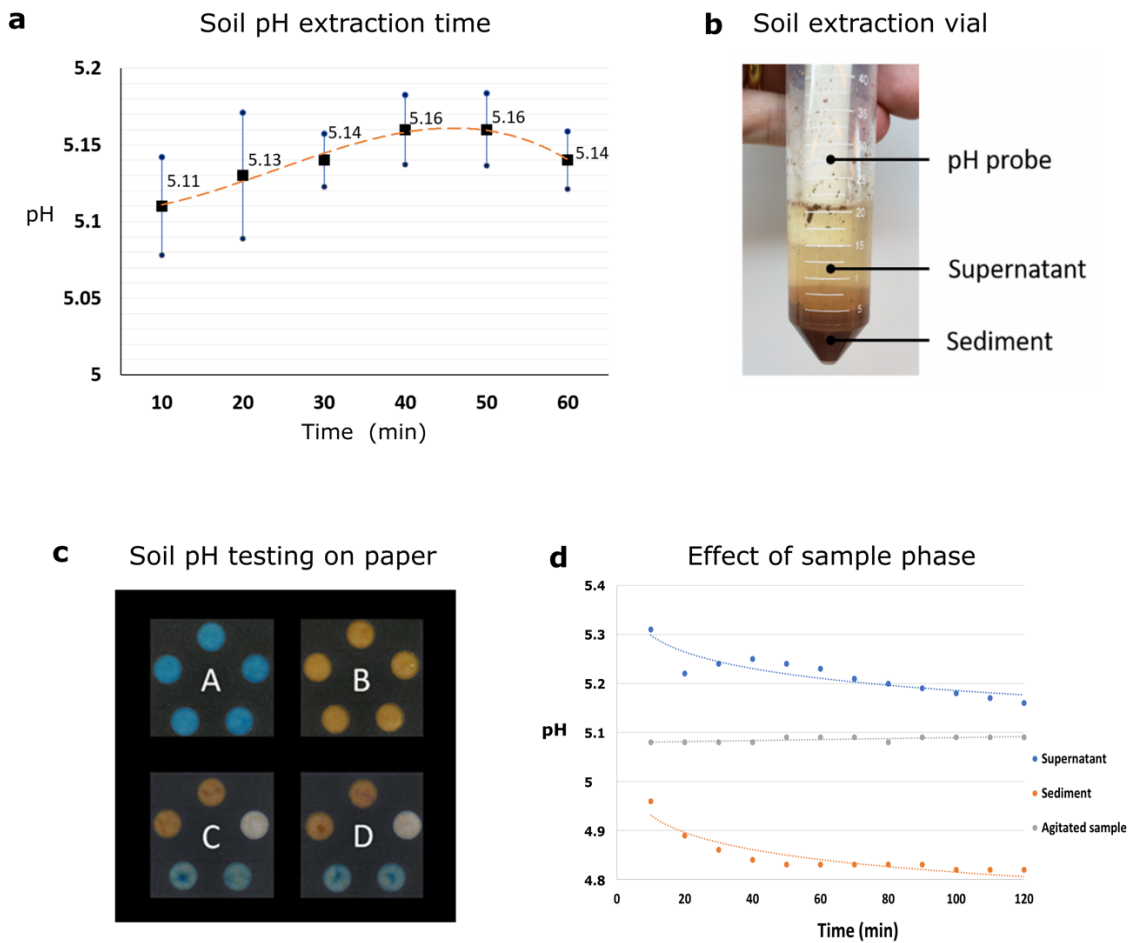

**Figure S6.** Soil sample extraction protocol. (a) Measured pH of soil extract as function of time. The dashed line is a guide to the eye. (b) Soil extract formation with a region of particulate matter at the bottom of the solution and a supernatant region on top. (c) Influence of the presence of particulate matter on colorimetric test results. (d) Effects of soil extract sampling and extraction time on measured soil pH. The solid lines are guides to the eye.

The use of the soil extraction protocol in combination with the paper-based devices requires further consideration. During extraction in the vial, the soil solution forms regions of particulate matter at the bottom and supernatant at the top, see Figure S6b. The particulate matter could potentially cause clogging of the flow pathways and influence the color formation in the paper device. This effect can be seen in Figure S6c where 40μL of extract was deposited on each paper input spot and allowed to react with the indicator for a few minutes until it produced a colorimetric output. In this example, extract from the supernatant region was deposited on devices A and B that were treated with reagents Bromocresol Green (BCG) and Bromocresol Purple (BCP) on all 5 spots, respectively, while extract from the sediment region was used on devices C and D, identically treated with reagent BCP on two of the spots and BCG on another two of the spots, respectively. As a result of the higher concentration of particulate in the sample collected in the sediment region of the extract, the colorimetric output of devices C and D show color stains in the test output areas. In contrast, devices A and B with sample from the supernatant region display a rather homogeneous color distribution.

To mitigate the above issues, we have added to the extraction protocol an additional step in which the sample is shaken after 20 minutes of extraction time, followed by an additional 5-minute wait time for the formation of the supernatant. A small amount of soil extract is then pipetted from the supernatant region onto the input region of the paper-based device. Measuring the effect of the region of collection of the sample on the pH value, plotted in Figure S6d, we observed that this procedure reduced the pH difference in the soil extract, while the use of the supernatant removed the negative interference of the particulate material in the test.

Based on [Teixeira17], the field test extraction protocol for use with the paper-based device was refined as follows:
1. Collect soil with a soil sampler or probe
2. Sieve the soil sample through a 2mm mesh sieve.
3. Mix 2.5mL of 0.01M $CaCl_2$ solution (pH=5.5) for every mL of sieved soil in a plastic vial with cap and close.
4. Shake the solution for 60 seconds and wait for 20 minutes.
5. Shake the solution again for a few seconds and wait 5 mins for the formation of the supernatant.
6. Collect a few μL of supernatant extract with a pipette and place it on the paper device.

Typically, about 25 mL of extract solution was produced during the extraction step per soil sample, while less than 100μL was used for performing pH analysis with the paper device.

### Site conditions, soil sample collection and test execution

As a test site, we selected a soybean farm in the city of Patrocinio Paulista, in the state of São Paulo (SP), 375 km from São Paulo, SP, Brazil, at 750-meter sea level altitude, see Figure 4a of the main manuscript. The most important crops in the region are coffee and sugar cane, however, soybeans and corn are regularly grown on the test site. Before the test period, the area was exposed to very dry weather conditions, accumulating only 5mm of rain over the course of 100 days. Ten days prior to test activity the test site underwent soil correction treatment with application of 1.5t of limestone per hectare. In addition, a fire occurring three days prior to the test period had severely impacted the site's corn vegetation.

The testing area comprised about 9 hectares, which we divided into 9 cells having a size of about 1 hectare each, as indicted by the white lines in Figure 4a. We subdivided each cell in nine sampling zones, as indicated by the yellow lines in Figure 4a of the main manuscript. We localized the sampling zones using GPS geo-tagging and marked them by a yellow flag. At each demarcation location, we inserted a soil probe about 20 cm deep into the ground to extract a plug of soil, see Figure 3a of the main manuscript. Overall, we collected a total of 81 soil samples from the sampling zones over the course of the field test duration of three days. For each soil sample in the sampling zones delineated by dashed line in Figure 4c-right of the main manuscript, we performed an average of ten paper-based soil measurements to evaluate test reproducibility.

We stored the soil samples in labeled plastic vials and processed them in place with the extraction protocol described in the Methods Section. Once the extraction step was completed, we collected about 10 µL supernatant from the soil extract with a pipette (Research Plus 3123000020, Eppendorf) and placed it on each circular input spot on the paper test card as shown in Figure S7a-top. After about 5 mins wait time, we acquired an image of the colorimetric sensor output located at the backside of the test by using the mobile app, as shown in Figure S7a-bottom. Within a few seconds, the mobile app displayed the pH classification of the soil and we stored the test data set on the phone. All measured data were streamed to the cloud computing database as soon as an internet connection became available.

## Benchmark against soil laboratory analysis

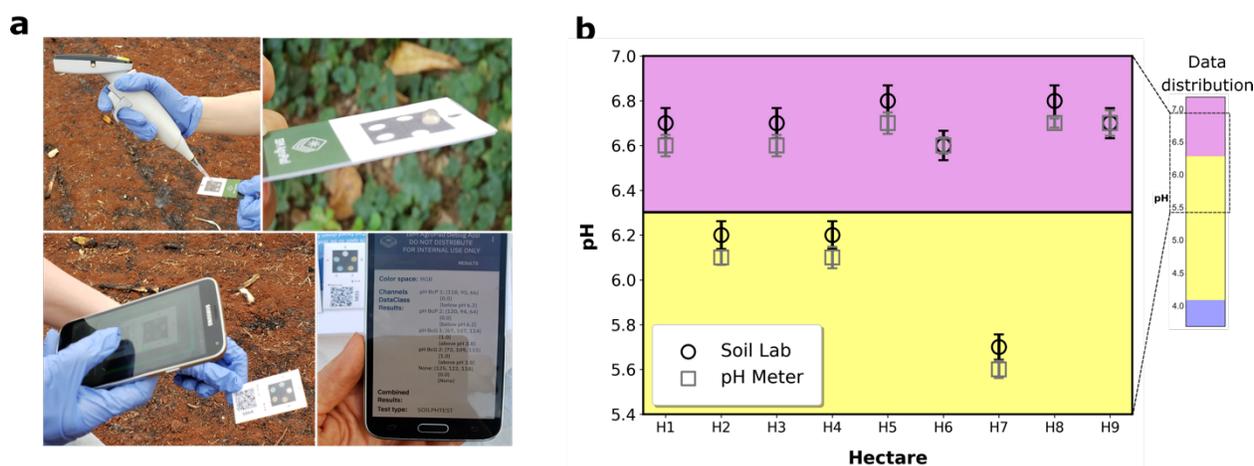

**Figure S7.** (a) Soil extract measurement with the paper-based test card and the mobile phone app. (b) Benchmarking of compound samples: pH-values retrieved from soil laboratory analysis and pH-meter analysis performed in our lab with the same extraction method applied in the field.

For establishing the reference data set, we measured all collected soil samples with a reference pH-meter (Simpla 140, AKSO) directly on site. For benchmarking purpose, we sent compound soil samples to a soil laboratory for routine analysis and to our own lab for pH-meter analysis.

As shown in Figure 4b of the main manuscript, we produced a compound soil sample for each hectare by combining the nine samples taken from each of the sampling sub-zones. One portion of each compound sample was then sent to the soil analysis lab and another portion was sent to our own laboratory for measurement with the reference pH-meter. Figure S7b plots the pH values retrieved from the soil lab and those obtained from the same compound samples by using the pH-meter in our lab, showing good agreement. The results confirm that the soil extraction protocol we have applied in our study yields pH results consistent with standard lab analysis outcomes.

## Analysis of paper-based pH measurements

The plots in Figure S8 show the distribution of the 805 paper-based test results with regards to soil pH as established by the reference measurement with the pH-meter. Accuracy is determined by whether the correct pH class was predicted with the paper-based test. For each paper sensor, two data points were extracted by processing the outcome of the four test spots, in pairs of one BCG and one BCP each, following the logic in Figure S5d. The results were then benchmarked against the pH-meter measurements. We binned the pH data in increments of 0.2 and within each bin, the green fraction represents the proportion of correct classification while the red represents the proportion of incorrect ones. Figure S8a displays the test measurement accuracy according to a visual interpretation of the colorimetric result on the paper card by an expert user and Figure S8b displays the results of the classification model retrieved from the mobile phone app. Overall, the binned results in Figure S8 demonstrate that the paper-based test system correctly predicted classifications consistent with the measurements by the pH-meter in over 70% of the cases. As perhaps understandably, the results show a reduced classification accuracy for samples with a pH-value close to the class boundary occurring at pH=6.3, in both visual inspection of the colorimetric response as well as automated mobile app classification. As discussed in the following, improving the classification accuracy at the class boundary is possible with further refinement and optimization of the measurement protocol.

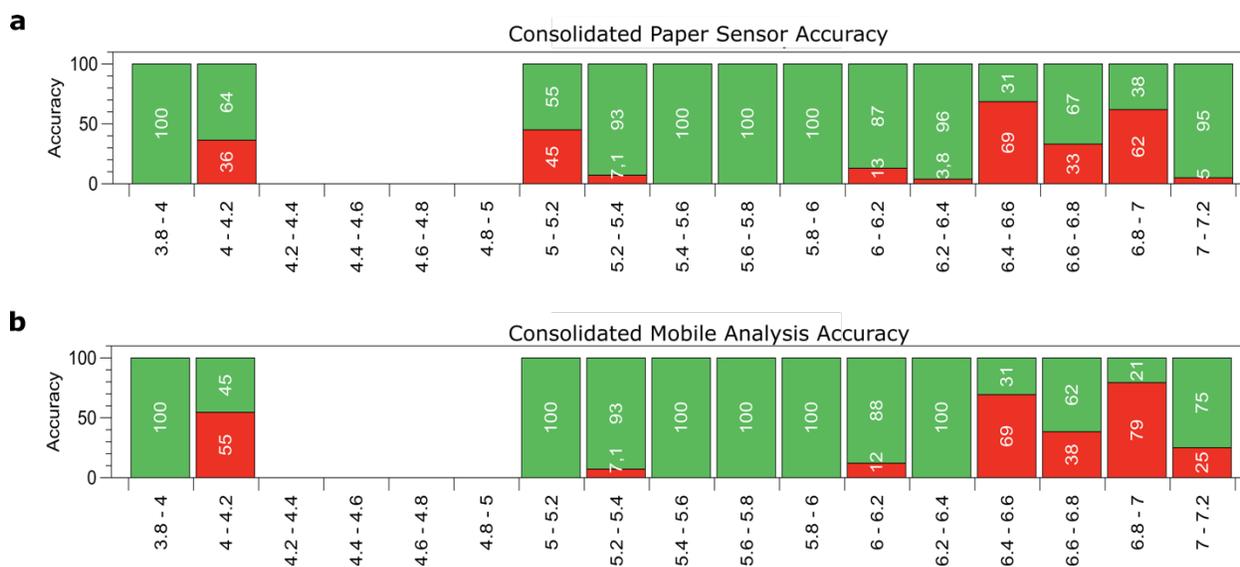

**Figure S8.** Distribution of paper-based tests results with regards to the pH-value of the soil sample as determined with the pH-meter. Green fractions represent the correct classifications, and the red fractions represents incorrect classifications. The test numbers of each fraction are indicated in white. (a) Measurement accuracy of visual interpretation of the colorimetric output on the test card by an expert user. (b) Measurement accuracy with model applied through mobile phone measurement.

By analyzing the colorimetric reaction dynamics of the BCP indicator within a test device (all spots BCP) under laboratory conditions, we observed that at the class boundary around pH= 6.3 the device requires significantly longer reaction times for producing the color output. Figure S9a displays the reaction dynamics. At pH=6.2, the time to achieve color output raises to above 125 seconds whereas it remains well below 50 seconds away from this value. The results indicate that the test accuracy at the class boundary could be improved by optimizing the wait time before measurements are performed.

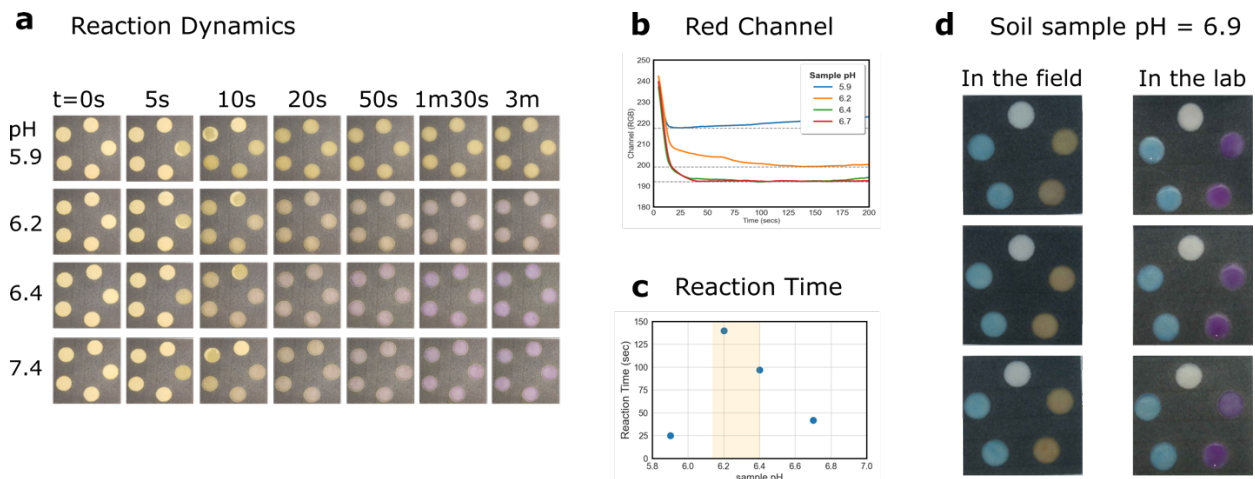

**Figure S9.** (a) Colorimetric reaction dynamics measured on four test paper devices where all the spots are treated with the BCP indicator exhibit temporal variations output color formation. (b) Time evolution of the camera recording in the red channel averaged from all five spots on each paper device. (c) Colorimetric reaction times extracted from the evolution of the red channel dynamics in (b). (d) Color formation with the same soil sample extract at pH=6.9 performed (left) in the field - suffering premature sample evaporation - and (right) in the lab - without premature sample evaporation.

In addition, we investigated potential miss-classifications in the pH range of 6.3-7.0 for field measurements that were impacted by the varying weather conditions experienced during the field test. Those measurements were expected to produce a clear colorimetric response; however, they did not develop well mainly due to high temperatures. The high temperature might have driven premature evaporation of the liquid soil extract sample prior to completion of the colorimetric reaction. The result of this effect is shown in Figure S9d. We conclude that both the volume of sample as well as the wait time should be adjusted to account for weather conditions, for instance by increasing the sample volume on warmer temperatures. We found that about 180 out of 615 measurements from 20 out of the 54 sampling sites analyzed could have potentially been impacted by the effects of premature sample evaporation. They included cases with the correct classification result, however, based on the date of collection those tests were deemed compromised. Based on our analysis, increasing the sample volume to 15μL and the reaction time to 120s should ensure proper test operation even at high temperatures.

To analyze the potential accuracy improvement with refined measurements conditions, we have replaced the compromised field measurements by a set of measurements repeated on the same soil samples in our lab and the accuracy results are plotted in Figure S10. As compared with the results shown in Figure S8, the corrected data set containing 548 test results clearly displays an improved classification accuracy at the class boundary occurring at pH=6.3, as well as an improvement in the range between 6.3 and 7.0 due to avoidance of premature sample evaporation.

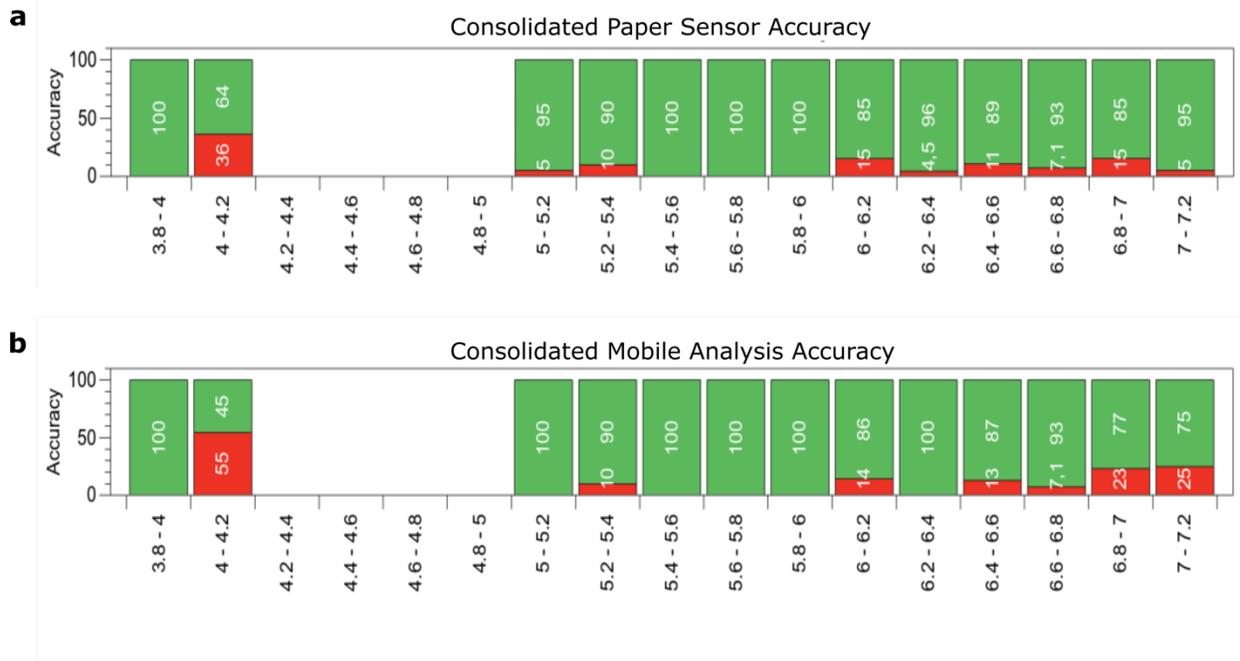

**Figure S10.** Distribution of paper-based tests results with regards to the pH-value of the soil sample as determined with the pH-meter. Green fractions represent the correct classifications, and the red fractions represents incorrect classifications. The test numbers of each fraction are indicated in white. As compared to the data set shown in Figure S8, about 180 field measurements compromised by weather conditions were replaced by laboratory repetitions as described in the text. (a) Measurement accuracy of visual interpretation of the colorimetric output on the test card by an expert user (b) Measurement accuracy with model applied through mobile phone measurement.

To compare the accuracy of paper-based test results as seen on the map of the farming area, we grouped the pH-meter results in three classes according to the logic laid out in Figure S5d, that is, "low soil pH" or class 0 for pH below 3.9, "medium soil pH" or class 1 for pH between 3.9 and 6.3, and "high soil pH" or class 2 for pH above 6.3. The spatial distribution of the pH-values as measured by the pH-meter based on that classification scheme is displayed in Figure S11a. As the pH of the soil varied between 5.5 and 7.0, only two of the three pH classes occur on the map. When compared to the results compound sample results in Figure 4, we observe a richer variation of pH due to the higher spatial resolution. Figure S11b displays the pH class distribution as measured by the paper tests for the dataset collected from the field. For each sample, we determined the pH class by the majority of ten paper test measurements results on the same soil sample, that is, a class 1 result means that 50% or more of the paper-based measurements had produced a class 1 outcome. As a result, we observe 13 misclassifications in 51 zones, corresponding to a classification accuracy of 75%. Figure S11c shows the paper-based test results after they were corrected for compromising weather conditions (premature sample evaporation) with repeat lab measurements performed on the same soil samples. As a result, in Figure S11d we obtain an improved classification accuracy with only three misclassifications in a total of 54 zones, boosting pH classification accuracy to 94%.

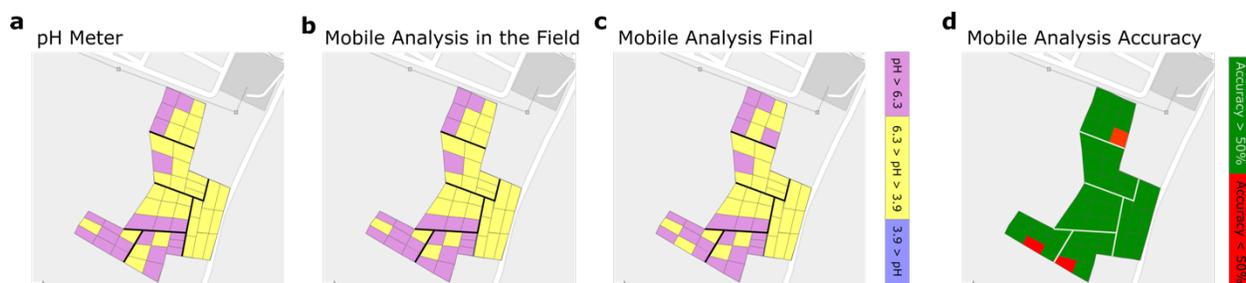

**Figure S11.** (a) Spatial distribution of the soil pH as measured by the pH-meter and grouped in three classes, that is, "low soil pH" for pH below 3.9, "medium soil pH" for pH between 3.9 and 6.3, and "high soil pH" for pH above 6.3. (b) pH class distribution as measured by the paper-based test for the dataset collected from the field. The final class per sampling site is determined per majority of test outcomes. (c) Corrected pH class distribution after replacing measurements compromised by weather conditions with lab measurements on the same soil samples. (d) Paper-based test accuracy map after replacing measurements compromised by weather conditions with lab measurements on the same soil samples.

## Multi-Parameter Test Prototype

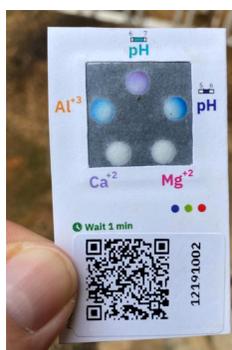

**Figure S12.** Paper-based sensor prototype integrating colorimetric indicators for the simultaneous detection of soil pH, Aluminum, Calcium and Magnesium ions.

## References

[Motsara08] Motsara, M. R., & Roy, R. N. (2008). *Guide to laboratory establishment for plant nutrient analysis* (Vol. 19). Rome: Food and Agriculture Organization of the United Nations.

[Hazelton16] Hazelton, P., & Murphy, B. (2016). *Interpreting soil test results: What do all the numbers mean?*. CSIRO Publishing.

[Teixeira17] Teixeira, P. C., Donagemma, G. K., Fontana, A., & Teixeira, W. G. (2017). *Manual de métodos de análise de solo*. Brasília: Embrapa, 573.